\documentclass[%
reprint,
superscriptaddress,
amsmath,amssymb,
aps,
prl,
longbibliography
]{revtex4-2}

\usepackage{graphicx}
\graphicspath{ {Figures/} }
\usepackage{comment}

\usepackage{dcolumn}
\usepackage{bm}
\usepackage[export]{adjustbox}
\usepackage[normalem]{ulem}

\usepackage{mathtools} 
\DeclarePairedDelimiter\bra{\langle}{\rvert}
\DeclarePairedDelimiter\ket{\lvert}{\rangle}
\DeclarePairedDelimiterX\braket[2]{\langle}{\rangle}{#1 \delimsize\vert #2}

\usepackage[utf8]{inputenc}
\usepackage[T1]{fontenc}

\usepackage{mathrsfs}
\usepackage{dsfont}

\usepackage{amssymb}
\usepackage{amsmath}
\usepackage{amsfonts}
\usepackage{amsthm}

\usepackage{hyperref}
\hypersetup{pdfpagemode=UseNone}

\newtheorem{theorem}{Theorem}

\newtheorem{lemma}[theorem]{Lemma}

\newcounter{rem}
\newcommand{\mc}[1]{\mathcal{#1}}

\def\>{\rangle}
\def\<{\langle}
\newcommand{\proj}[1]{| #1 \rangle\! \langle #1 |}
\newcommand{\idty}{\mathds{1}}
\def\tr{{\rm tr}}
\DeclareMathOperator*{\pr}{Pr}
\DeclareMathOperator*{\SpanOp}{span}

\def\ii{{\rm i}}
\def\rho{{\varrho}}

\def\textbf#1{{\bf #1}}
\newcommand{\Cx}{\mathbb{C}}

\gdef\lvert{\delimiter"426A30C }
\gdef\rvert{\delimiter"526A30C }
\gdef\lVert{\delimiter"426B30D }
\gdef\rVert{\delimiter"526B30D }
\newcommand{\norm}[1]{\lVert#1\rVert}
\newcommand{\Norm}[1]{\big\lVert#1\big\rVert}

\usepackage{xcolor}

\begin{document}

\title{A finite-resources description of a measurement process\\ 
and its implications for the ``Wigner's Friend'' scenario
}

\author{Fernando de Melo}
\email{fmelo@cbpf.br}
\affiliation{Centro Brasileiro de Pesquisas F\'{\i}sicas, Rua Dr. Xavier Sigaud, 150, Rio de Janeiro, RJ, Brazil}
\author{Gabriel Dias Carvalho}
\affiliation{Física de Materiais, Escola Politécnica de Pernambuco, Universidade de Pernambuco, 50720-001, Recife, PE, Brazil}
\author{Pedro S. Correia}
\affiliation{Departamento de Ciências Exatas, Universidade Estadual de Santa Cruz, Ilhéus, Bahia 45662-900, Brazil}
\author{Paola Concha Obando}
\affiliation{School of Physics, University of the Witwatersrand, Private Bag 3, Wits 2050, South Africa}
\author{Thiago R. de Oliveira}
\affiliation{Instituto de Física, Universidade Federal Fluminense, Av. Litorânea s/n, Gragoatá 24210-346, Niterói, RJ, Brazil}
\author{Ra\'ul O. Vallejos}
\affiliation{Centro Brasileiro de Pesquisas F\'{\i}sicas, Rua Dr. Xavier Sigaud, 150, Rio de Janeiro, RJ, Brazil}
\date{\today}

\begin{abstract} Quantum mechanics started out as a theory to describe the smallest scales of energy in Nature. After a hundred years of development it is now routinely employed to describe, among others, quantum computers with thousands of qubits. This tremendous progress turns the debate of  foundational questions into a technological imperative. In what follows we introduce a model of a quantum measurement process that consistently includes the impact of having access only to finite resources when describing a macroscopic system, like a measurement apparatus. Leveraging  modern tools from equilibration of closed systems and typicality, we show how the measurement collapse can be seen as an effective description of a closed dynamics, of which we do not know all its details. Our model is then exploited to address the ``Wigner Friend Scenario'', and we observe that an agreement is reached when both Wigner and his friend  acknowledge their finite resources perspective  and describe the measurement process accordingly.
\end{abstract}
\maketitle

\textit{Introduction.} 
Quantum mechanics is one of the most successful theories of science. In the same breadth, it precisely describes the most fundamental constituents of Nature \cite{feynmanV3}, and it also enables the development of technologies with tremendous impact on human society \cite{richards2012budget}. While initially the theory mostly dealt with the smallest scales of atomic levels \cite{feynmanV3}, nowadays, due to the exquisite experimental control \cite{electrong, deleglise2008}, it is possible to observe quantum phenomena in various complex systems\cite{Eibenberger,haochun22, google}, which reinforce the universality of quantum mechanics.

Although evermore persuasive, such universality puts a spot in one of the most difficult issues of quantum mechanics, the quantum-to-classical transition or, more specifically, on the quantum measurement problem \cite{brukner2017}. Within the postulates of quantum mechanics there are two types of dynamics: The first type involves the intrinsic, continuous evolution of isolated quantum systems, governed by Schrödinger’s equation. The second type concerns the measurement process, an interaction between a quantum system and a measurement apparatus, which leads to the ``collapse'' of quantum superpositions and to a measurement outcome. For this second type of process there is no dynamical equation, and the measurement apparatus is assumed classical. The issue is then: if quantum mechanics is universal, why should it resort to a classical apparatus in its very foundations? If it is not universal, when should one use one type of dynamic or another? 

Such a conundrum is most vividly illustrated by the so-called ``Wigner's Friend Scenario''~\cite{Wigner1967}, which is commonly described as follows. Suppose that Wigner runs a lab where a spin is being measured in the $\{\ket{+},\ket{-}\}$  basis by a friend of his. The measurement apparatus starts in a ``ready'' state denoted by $\ket{0}$. Assume that the apparatus is calibrated such that if the spin is prepared in state $\ket{+}$ ($\ket{-}$), after the measurement the apparatus' pointer will be in the $\ket{+1}$ ($\ket{-1}$) state, with the apparatus' states classically distinguishable.  Now suppose that the spin is prepared in the state $\ket{\psi}=(\ket{+}+\ket{-})/\sqrt{2}$. In such a case, the postulate of quantum mechanics describing this measurement process entails that fifty percent of the time, the spin and apparatus will end up in the state $\ket{+}\otimes\ket{+1}$, while in the other fifty percent of instances the final state is $\ket{-}\otimes\ket{-1}$. After many rounds, the statistical description that Wigner's friend arrives at for the system after measurement  is $(\proj{+,+1}+\proj{-,-1})/2$  (where we used standard shortened notation).

At this point, we turn our attention to the description by Wigner, who is outside of the lab and assumes the universality of quantum mechanics. As such, he describes both the spin and the measurement apparatus (even possibly his friend in the lab) as quantum mechanical systems which undergo unitary dynamics -- a description of the measurement process championed by von Neumann~\cite{vonNeumann1955}. If the spin and apparatus dynamics seen by the friend is now described by a unitary  $U$, it has to be such that $U\ket{+}\otimes \ket{0}= \ket{+}\otimes \ket{+1}$, and  $U\ket{-}\otimes \ket{0}= \ket{-}\otimes \ket{-1}$. By the linearity of quantum mechanics
$U\left(\frac{\ket{+}+\ket{-}}{\sqrt{2}}\otimes \ket{0}\right)= \frac{\ket{+,+1}+\ket{-, -1}}{\sqrt{2}}$,
which correlates spin and apparatus in the expected way, but does not lead to a mixture of definite outcomes.
In Wigner's perspective, the state after every run of the measurement process is the entangled state $\ket{\phi^+}:= (\ket{+,+1}+\ket{-, -1})/\sqrt{2}$. 

The fact that Wigner and his friend come to different descriptions is not an issue for most modern interpretations of quantum mechanics~\cite{schmid2023review}. However, following Deutsch's argument~\cite{Deutsch1985}, suppose that in each round of the experiment, after receiving from his friend the information that a measurement took place, Wigner measures both the spin and the measurement apparatus with a two-valued observable with associated projectors $\{ \proj{\phi^+}, \idty - \proj{\phi^+}\}$.  If one uses Wigner's description, the system is projected in $\ket{\phi^+}$ with certainty; while if using the Friend's description that happens only half of the times. This mismatch of assigned probabilities for a fixed experimental scenario is a legitimate issue for most interpretations of quantum mechanics.

In the above recount of the Wigner's Friend scenario, it was assumed, as usual, that Wigner does have the capacity to perform a projective measurement in the combined system of a spin and measurement apparatus, even when the latter is a macroscopic system that supposedly behaves classically. At this point most authors mention that performing such a measurement is a major challenge, but they go on to say that it can be done ``in principle''. 

In Physical sciences, nevertheless, such ``in principle'' statements cannot disregard  the bare fact that only finite resources are available to implement a physical process. One striking example where accounting for finite resources clears up an apparent paradox is the case of Maxwell's demon. When the finiteness of the demon's memory is taken into account, one reaches the physically sounding conclusion that no work can be extracted from heat~\cite{bennett1982thermodynamics}. 

This mindset of thermodynamics can be exploited to shine a light on the measurement process~\cite{viennaGuys,engineer2024}, and thus also on the Wigner's Friend scenario. Such a connection can be made by realizing that the capacity of performing a joint measure in the spin and the measurement apparatus is equivalent to being able to reverse the unitary dynamics characterizing the measurement process~\cite{schmid2023review}. In Wigner's perspective, if the measurement process is undone, the whole system returns to the initial state $(\ket{+}+\ket{-})\otimes\ket{0}/\sqrt{2}$. While in the Friend's perspective, half of the time, the system returns to $\ket{+}\otimes\ket{0}$, and in the other half, it goes to $\ket{-}\otimes\ket{0}$. Therefore, by measuring only the spin in the basis $\{(\ket{+}+\ket{-})/\sqrt{2},(\ket{+}-\ket{-})/\sqrt{2}\}$, which is routinely performed in labs all over the world, one reaches the same dilemma.

This latter narrative of the Wigner's Friend scenario alludes to the (ir)reversibility of macroscopic systems, a subject much debated in the foundations of statistical physics~\cite{lebowitz2007time,brown09}. As in there, in what follows, we are going to argue that no paradox is reached in the Wigner's Friend scenario when a description of the measurement process that uses only finite resources is employed.  When a process that can be called a quantum measurement takes place, it is in fact irreversible. 

A by now standard way to describe irreversible processes within quantum mechanics is by the open quantum system formalism~\cite{breuer2002}, with the so-called decoherence program aiming to address the quantum-to-classical transition~\cite{zurekRMP}. However, this program is known to not solve the measurement problem~\cite{maudlin1995three, decoherence_RMP}.  Differently from the decoherence program, here we are going to account for the finiteness of resources at the level of the Hamiltonian describing the system's dynamics, and will treat the measurement process as the effective equilibration of a closed complex quantum system.

\textit{A finite-resource model of quantum measurement.}
Here, we introduce a general model describing the spin measurement employing only finite resources. For other models see for instance~\cite{schlosshauer, allahverdyan2013, gabriel2020, pranzini2024}.

To the spin we associate a Hilbert space $\mc{H}_S\simeq \Cx^2$, spanned by the vectors $\{\ket{+},\ket{-}\}$. The initial state of the spin system will be $\ket{\psi}=c_+\ket{+}+c_-\ket{-}$, with $c_+,c_-\in \Cx$ such that $|c_+|^2+|c_-|^2=1$. We write $\psi=\proj{\psi}$ for its density matrix.

For the apparatus we associate a $d$-dimensional Hilbert space $\mc{H}_A\simeq \Cx^d$, which we write as a direct sum of three subspaces, $\mc{H}_A = \mc{H}_{+1}\oplus \mc{H}_{0} \oplus \mc{H}_{-1}$. The first subspace is to be associated with the outcome $+1$, the second with the outcome $0$, and the third with the outcome $-1$.  Without loss of generality we assume that $\dim{\mc{H}_{+1}}=\dim{\mc{H}_{-1}}:=d_1$, and define $\dim{\mc{H}_{0}}:=d_0$, leading thus to $d=2d_1+d_0$. The space $\mc{H}_0$ relates to the apparatus' initial value. Since the only information we have is that the apparatus starts in the ``$0$'' value, its initial state will be taken as the microcanonical ensemble in the corresponding subspace, i.e., $\idty_0/d_0$. The total initial state then reads:   $\rho_0=\psi\otimes\left(0\oplus \frac{\idty_0}{d_0}\oplus 0\right)$.

We can now define the model's Hamiltonian:
\begin{equation}
H= \proj{+}\otimes (H_{[+1,0]}\oplus 0)+\proj{-}\otimes (0\oplus H_{[0,-1]}),
\label{eq:hamiltonian}
\end{equation}
where $H_{[+1,0]}$  has support only in $\mc{H}_{+1}\oplus \mc{H}_0$, and $H_{[0,-1]}$  has support only in $\mc{H}_{0}\oplus \mc{H}_{-1}$.  The rationale behind this Hamiltonian is simple.  With the apparatus starting in a state  with support only on $\mc{H}_0$, any component of the spin system along $\ket{+}$ will induce transitions from $\mc{H}_0$ to $\mc{H}_{+1}$ and back. 
Similarly, any component of the spin system along $\ket{-}$ will induce transitions between $\mc{H}_0$ and $\mc{H}_{-1}$. If $\lim_{d\rightarrow \infty} d_0/d_1=0$ we expect that any initial population in $\mc{H}_0$ will be proportionally  transferred to $\mc{H}_{+1}$ or $\mc{H}_{-1}$, and there it will remain for a long time~\footnote{To be able to experimentally prepare the apparatus' initial state we must require $d_0/d_1\rightarrow 0$ as $\textrm{poly}(1/d)$.}. Additionally, as we are describing the dynamics of a macroscopic system, we do not assume to know $H$ with infinite precision~\cite{santo2019,gisin2020}. There is always an experimental fluctuation associated with the elements of the Hamiltonian.  Thus the very description of the evolved state is not known, and one has to average over the Hamiltonian's uncertainty. To consistently include this uncertainty, our results will not depend on details of $H$. The only assumption on the Hamiltonian, besides its structure as in \eqref{eq:hamiltonian}, will be that both $H_{[+1,0]}$ and $H_{[0,-1]}$ are non-degenerate and have non-degenerate gaps.



Below we follow~\cite{maudlin1995three} and discuss three key aspects of a measurement: the assigned probabilities, its dynamics, and the  after measurement states. 

\textit{i - Measurement Probabilities: Equilibrium value.}
Here, we understand the quantum measurement as an effective  equilibration process.  A given observable is said to be in equilibrium if its expectation value for almost all times does not differ substantially from its equilibrium value. The equilibration of closed complex quantum systems is a much debated subject these days~\cite{tasaki98,reimann2008,linden09,gogolin2016}, and the connection between the measuring process and equilibration dynamics has been made before with a recent renewed interest~\cite{viennaGuys, engineer2024}.

The key result that we are going to employ is the following one. 
Given a time-independent Hamiltonian $H$, its equilibrium state  $\rho_\infty^H$ is defined as the time-averaged state  $\< \rho_t^H \>_\infty= \lim_{T\rightarrow \infty}\frac{1}{T}\int_0^T dt  \rho_t^H $, 
where $\rho_t^H := e^{-\ii H t} \rho_0 e^{\ii H t}$ is the evolved state of the system, and $\< \cdot \>_\infty$ represents the infinite time average, and $\hbar$ is set to unity. If the Hamiltonian is not degenerate and has no degenerate gaps, it has been shown in~\cite{tasaki98,reimann2008,linden09}, 
and further developed by many authors~\cite{Short2011,Short2012,gogolin2016}, that for any observable $A$:
\begin{equation}
   \<|\tr(A \rho_t^H ) - \tr(A \rho^H_\infty)|^2\>_\infty\le \frac{\norm{A}^2}{d^H_\text{eff}(\rho_0)}.
   \label{eq:equilibration_bound}
\end{equation}
In the expression above, $\norm{\cdot}$ is the operator norm -- the largest singular value --, and $d^H_\text{eff}(\rho_0)$ is the effective dimension of the initial state in the energy eigenbasis $\{\ket{E_i}\}$ of Hamiltonian $H$, i.e.,
$d^H_\text{eff}(\rho_0):=1/(\sum_i \<E_i|\rho_0|E_i\>^2)$. The equilibration bound~\eqref{eq:equilibration_bound}  implies that most of the times the expectation value of $A$ is no further from its equilibrium value by more than $\norm{A}/\sqrt{d^H_\text{eff}(\rho_0)}$. If this latter quantity is small, compared to the measurement resolution, then property $A$ will be deemed effectively equilibrated.

Returning to our model, described by the Hamiltonian \eqref{eq:hamiltonian} and initial state $\rho_0$, we can now analyze the equilibration process for the probabilities of finding the values $+1$, $0$, and $-1$ for the apparatus. In order to not depend on fine details of the Hamiltonian, but still to abide by the conditions of bound~\eqref{eq:equilibration_bound}, we take $H$, concretely $H_{[+1,0]}$ and $H_{[0,-1]}$, as fixed instances from the Gaussian Unitary Ensemble (GUE) -- a well-studied ensemble in random matrix theory that preserves universal features~\cite{Forrester2010}. As the probability of obtaining the outcome $i\in\{+1,0,-1\}$ at time $t$ is given by $\pr(i|H)_t=\tr(\rho_t^{H} \idty_S\otimes\Pi_i)$, with $\Pi_i$ the projector onto the subspace $\mc{H}_i$ of $\mc{H}_A$, we employ  bound~\eqref{eq:equilibration_bound} for the projectors to obtain, for large dimension $d$:%
\begin{equation}
\label{eq:eq_probs}
    \<|\pr(i|H)_t- \pr(i|H)_\infty |^2\>_\infty\le \frac{1}{d_0+d_1}, 
\end{equation}
with $\pr(i|H)_\infty:= \tr(\rho^H_\infty \idty_S\otimes\Pi_i)$. Crucial to obtain this result was that for any dimension $\norm{\Pi_i}=1$, i.e., the greater the apparatus' dimension is, the more coarse-grained is  the observable being measured~\cite{cris2017,vallejos2022,pedrinho_PRL24}.  Details of this derivation are shown in~\cite{SM}. Therefore, as the dimension of the apparatus increases, the closer are the probabilities from their equilibrium values for most times.

The equilibrium value, nevertheless, can be different for each choice of $H$. But since we do not assume full knowledge about $H$, we employ Chebyshev's inequality and Haar averages~\cite{Mele2024, RTNI} to show that, for large apparatuses, all typical choices of $H$ lead, with high probability, to the same equilibrium value:
\begin{equation}
\label{eq:chebyshev_probs}
    \pr_{H\sim \text{GUE}} \left( \left| \pr(i|H)_\infty - \overline{\pr(i|H)_\infty} \right| \ge \delta \right) \le \frac{4  (d_0+1)}{\delta^2 d_0 d_1^3}.
\end{equation}
In the above expression $\delta>0$, and the over-bar represents the average over $H$, i.e., with $H_{[+1,0]}$  and $H_{[0,-1]}$ taken uniformly and independently from the GUE ensemble.  As the size of the apparatus grows ($d_0$ and $d_1$ increase with $\lim_{d\rightarrow\infty} d_0/d_1=0$), the  probability of small deviations of $\pr(i|H)_\infty$ from its average equilibrium value becomes negligible. In other words, the probabilities of measuring $+1$, $0$ and $-1$, for large apparatuses, are insensitive to fine details of the Hamiltonian and are close to their typical value.

Finally we can explicitly evaluate the typical equilibrium probabilities~\cite{Meckes2019,Mele2024} (see~\cite{SM} for details):
\begin{equation}
\begin{aligned}
\overline{\pr(\pm 1|H)_\infty} &= |c_\pm |^2\frac{d_1}{d_1+d_0+1};\\
\overline{\pr(0|H)_\infty} &= \frac{d_0+1}{d_1+d_0+1}.
\end{aligned}
\label{eq:typical_probs}
\end{equation}
Chaining results~\eqref{eq:eq_probs},~\eqref{eq:chebyshev_probs}, and~\eqref{eq:typical_probs}, we have shown that, for any typical choice of $H$, as the apparatus' size grows --  $\lim_{d\rightarrow \infty} d_0/d_1= 0$  -- the probabilities of measuring $\pm 1$ equilibrate, in any run,  to their text-book values of $|c_\pm|^2$, and the probability of measuring $0$ becomes vanishingly small.

\textit{ii - Measurement dynamics: state equilibration.}
Now we describe how the measurement dynamics actually takes place. 
We are interested in the average equilibration of the total system description, i.e., we want to show that, 
for most of the times, $\rho_t^H$ is basically indistinguishable from  $\rho_\infty^H$.  Although we are not concerned with the equilibration of a single expectation value, as in the case of probabilities, we grant that  measuring an exponentially large set of observables is not feasible for a macroscopic system.

Following~\cite{Short2011}, assume that in trying to distinguish $\rho_t$ from  $\rho_\infty^H$ one has access to a set $\mc{M}$ of general measurement operators (positive operator valued measure, POVM). Given this set, the distinguishability between the states is defined as $\mc{D}_\mc{M}(\rho_t^H,\rho_\infty^H):=\max_{M\in\mc{M}}\sum_r|\tr(M_r \rho_t^H)- \tr(M_r \rho_\infty^H)|/2$, with the summation over the outcomes of a POVM $M\in \mc{M}$. As a corollary of bound~\eqref{eq:equilibration_bound}, for large dimensions, we obtain (details in~\cite{SM}):
\begin{equation}
    \<\mc{D}_\mc{M}(\rho_t^H,\rho_\infty^H)\>_\infty\le \frac{\mc{S}(\mc{M})}{4\sqrt{d_0+d_1}},
    \label{eq:bound_dist}
\end{equation}
where $\mc{S}(\mc{M})$ is the total number of outcomes for all POVMs in $\mc{M}$. 
In this way, if $\mc{S}(\mc{M})$ does not grow faster than $\sqrt{d_0+d_1}$ 
-- which typically increases exponentially with the number of particles constituting the apparatus --, the states are effectively indistinguishable, i.e., the system is in equilibrium. Here, again, we see that accounting for resources that do not grow in a unphysical manner,  leads to an effective equilibration.

Once more, as we do not assume full knowledge of the Hamiltonian $H$, there is no way to explicitly write $\rho_t^H$ or $\rho_\infty^H$. 
As we did for the probabilities earlier, now we establish that for macroscopic apparatuses $\rho_\infty^H$ is, for any choice of $H$, close to the typical average equilibrium state $\overline{\rho_\infty}$ -- where again the average is performed over the Hamiltonians sampled uniformly from GUE. After~\cite{popescu2006,pedrinho_PRL24}, this is done in two parts (details in~\cite{SM}). First we verify that the average distance between these states decreases as the apparatus' dimension increases, as demonstrated by the bound:
\begin{equation}
    \sigma:=\overline{\Norm{\rho_\infty^H-\overline{\rho_\infty}}_2^2}\le \frac{d_1}{d_0 (1 + d_0 + d_1)^2}\simeq\frac{1}{d_0d_1}+ \mc{O}(d_1^{-2}),
    \label{eq:bound_dist_sigma}
\end{equation}
where $\norm{A}_2=\sqrt{\tr(A^\dagger A)}$.
Second, employing Levy's lemma~\cite{low2009}, we  conclude that for almost all   choices of $H$  the distance between its equilibrium state and the typical equilibrium state is not far from the average distance. Concretely, for all $\delta>0$:
\begin{equation}
    \label{eq:levy_dist}
    \pr_{H\sim \text{GUE}} \left( \left|\Norm{\rho_\infty^H-\overline{\rho_\infty}}_2^2 -\sigma \right| \ge \delta \right) \le 8 \; e^{- c d_0^2 \delta^2},
\end{equation}
with  $c$ a  positive constant.
Bringing results~\eqref{eq:distinguishability},~\eqref{eq:bound_dist},~\eqref{eq:levy_dist} together, we've shown that for increasing apparatus' dimension, the description of the system is, for all practical purposes, most of the times indistiguishable from the typical equilibrium state.

Notably, the typical equilibrium state can be analytically evaluated (see Supplemental Material \cite{SM}), and reads:
\small
\begin{equation}
    \label{eq:fully_avg_state}
    \begin{split}
    \overline{\rho_\infty}=\frac{d_1}{d_0+d_1+1}|c_+|^2\proj{+}\otimes  \left(\frac{\idty_{+1}}{d_1}\oplus 0 \oplus 0\right)+\\
    +\frac{d_1}{d_0+d_1+1}|c_-|^2\proj{-}\otimes \left(0\oplus 0 \oplus \frac{\idty_{-1}}{d_1}\right)+ \\ +\frac{d_0+1}{d_0+d_1+1}\left(|c_+|^2\proj{+}+|c_-|^2\proj{-}\right)\otimes \left(0\oplus \frac{\idty_{0}}{d_0} \oplus 0\right).
    \end{split}
\end{equation}
\normalsize

Various comments about this description are now in place. First, note that it is a mixture of definite outcomes, as denoted by the classically distinguishable states in the apparatus. This is the major advantage of our model when compared to the traditional decoherence approach. The first term of \eqref{eq:fully_avg_state} is related to the outcome $+1$, the second to the $-1$, and the third to the $0$ outcome. Given that the details of the Hamiltonian are not known (and largely not important), and that this is consistently taken into account in our dynamical description, this equilibrium state fully characterizes the measurement completion.  Second, as this density matrix simply represents the mixture of classical possibilities, the probabilities can be directly read out from~\eqref{eq:fully_avg_state}. Notice that we only performed averages to obtain the above state and did not use Born's rule. One could say that, within our model, Born's rule dynamically emerges in the measurement process. Lastly,  as the size of the apparatus increases,  with the ratio $d_0/d_1$ going to zero, the above description approaches: 
\begin{multline}
    \lim_{N\rightarrow \infty}\overline{\rho_\infty}=|c_+|^2\proj{+}\otimes \left(\frac{\idty_{+1}}{d_1}\oplus 0 \oplus 0\right)+\\
+|c_-|^2\proj{-}\otimes \left(0\oplus 0 \oplus \frac{\idty_{-1}}{d_1}\right),
\label{eq:limit_state}
\end{multline}
which is the expected state from von Neumann's infinite-resource description of a measurement process. Notice that when the information about an outcome is known, that does not specify the microscopic state of the apparatus. As such, for each outcome, the corresponding description of the apparatus is the microcanonical ensemble in the corresponding subspace.


\textit{iii - Post-measurement state: approximate stationary states.}
From \eqref{eq:fully_avg_state}, as it represents a distribution over classically distinguishable states, it is immediate to apprehend what is the state of the system after a given measurement outcome. See Table~\ref{tab:post}.
\begin{table}[ht]
    \centering
    \begin{tabular}{ccc}\hline\hline
      \footnotesize event   &  \footnotesize probability& \footnotesize post-measurement state\\\hline\hline
       \footnotesize +1  & \footnotesize$\frac{d_1}{d_0+d_1+1}|c_+|^2$&\footnotesize$\proj{+}\otimes \left(\frac{\idty_{+1}}{d_1}\oplus 0 \oplus 0\right)$\\
       \footnotesize -1  &\footnotesize $\frac{d_1}{d_0+d_1+1}|c_-|^2$&\footnotesize$\proj{-}\otimes \left(0\oplus 0 \oplus \frac{\idty_{- 1}}{d_1}\right)$\\
       \footnotesize 0  &\footnotesize$\frac{d_0+1}{d_0+d_1+1}$&\footnotesize $\left(|c_+|^2\proj{+}+|c_-|^2\proj{-}\right)\otimes \left(0\oplus \frac{\idty_{0}}{d_0} \oplus 0\right)$\\\hline\hline
    \end{tabular}
    \caption{\small Post-measurement states for each outcome and their corresponding probabilities.}
    \label{tab:post}
\end{table}
Observe that in our model the post-measurement states are all orthogonal to each other, with the possibility of an erroneous result encoded in the outcome $0$. As previously discussed, the probability of this outcome approaches zero as the size of the apparatus increases with $d_0/d_1\rightarrow 0$. Furthermore, it should be stressed that in each round, the system does not stop to evolve. What it is shown is that for most of  times the system is  indistinguishable from  $\overline{\rho_\infty}$, Eq.~\eqref{eq:fully_avg_state}. In this respect, the description of the system is effectively stationary. Such approximate stationarity is also observed for the post-measurement states when an update in the system description must be performed. For the $+1$ outcome, only the $H_{[+1,0]}$ part of the Hamiltonian $H$ will generate dynamics, being restricted to the $\mc{H}_{+1}\oplus\mc{H}_{0}$ subspace. If $d$ is large enough, with $d_1\gg d_0$, the system stays basically frozen in $\mc{H}_{+1}$. The argument is similar for the $-1$ outcome. For the $0$ outcome, when the stationarity condition is observed, i.e., when $d_1\gg d_0$, then the probability of measuring $0$ gets vanishingly small.

Moreover, this effective stationarity is immediately connected to the irreversibility of the measurement process, viewed here as an equilibration process. Given that there is an intrinsic ignorance about the Hamiltonian driving the system, and that only a finite number of measurements on the system are possible to be performed, the description given by  $\overline{\rho_\infty}$, Eq.~\eqref{eq:fully_avg_state}, is the only possible one. After the measurement is performed, i.e., one of the classical outcomes actually happens, the corresponding post-measurement quasi-stationary states becomes the best description of the system.  In other words, it is not possible to exactly reverse a unknown dynamics.  Like in the classical case, as argued by Boltzmann in response to Loschmidt's challenge~\cite{darrigol2021boltzmann,binder2023reversibility}, the exact reversible of all microscopic degrees of freedom of a macroscopic system is, in principle, impossible. 

\textit{Back to Wigner and his friend.} The main point is that if Wigner and his friend are indeed describing a measuring process, then both of them cannot fully account, nor even in principle, for all the apparatus' degrees of freedom. Within such a perspective, using the collapse postulate or the unitary dynamics postulate, both will reach, for large apparatuses, the same description, namely the one given in Eq.~\eqref{eq:limit_state}.

\textit{Conclusions.}
In the above, we embraced a description of the measurement process that adequately incorporates the finiteness of resources in its essence. Differently from the approach taken by the decoherence program~\cite{zurekRMP,decoherence_RMP}, our model acknowledges that the Hamiltonian describing the dynamics of macroscopic systems cannot be known with certainty, and that the measurement being performed are highly coarse-grained. In a macroscopic system, with $\mathcal{O}(10^{23})$ degrees of freedom, a small energy uncertainty of $10^{-10^{22}}$ Joules  still contains an order of $10^{10^{23}}$ energy levels~\cite{Reimann_2010}. Such mind-boggling density of energy levels cannot be dismissed by physical theories aiming to describe macroscopic systems. 

As we showed, when the finiteness of resources is incorporated into the description of the measurement process, the universality of quantum mechanics may be regained, as it is possible to describe dynamics in all scales in the same way. In this viewpoint, the collapse postulate turns into a very practical effective description of a complex dynamics for which we do not have access to all its details. 
Moreover, this perspective also helps clear up the discussions revolving around the Wigner Friend Scenario, with Wigner and his Friend now reaching the same portrayal. Our finite-resources model of a quantum measurement can be easily applied to other seemingly paradoxical situations, for instance, the extended Wigner's Friend scenario~\cite{brukner2017,bong2020,delsanto2024}, and the 'Frauchiger and Renner' scenario~\cite{frauchiger2018}. 
 
\textit{Acknowledgements.}
We would like to thank Jacques L. Pienaar for instructive conversations. F.~d-M.~also thank Ruynet M. Filho and Marcelo F.~Santos for countless discussions at the H.P. DNA. 
This work is supported in part by the National Council for Scientific and Technological Development, 
CNPq Brazil (projects: Universal Grant No. 406499/2021-7, and 409611/2022-0), 
the Carlos Chagas Foundation for Research Support of the State of Rio de Janeiro (FAPERJ, Grant APQ1 E-26/210.576/2024),
and it is part of the Brazilian National Institute for Quantum Information. 
TRO acknowledges funding from the Air Force Office of Scientific Research under Grant No. FA9550-23-1-0092. Paola Obando acknowledges funding from South African Quantum Technology Initiative (SA QuTI) and the National Research Foundation (South Africa).

\bibliography{ref}

\begin{thebibliography}{50}%
\makeatletter
\providecommand \@ifxundefined [1]{%
 \@ifx{#1\undefined}
}%
\providecommand \@ifnum [1]{%
 \ifnum #1\expandafter \@firstoftwo
 \else \expandafter \@secondoftwo
 \fi
}%
\providecommand \@ifx [1]{%
 \ifx #1\expandafter \@firstoftwo
 \else \expandafter \@secondoftwo
 \fi
}%
\providecommand \natexlab [1]{#1}%
\providecommand \enquote  [1]{``#1''}%
\providecommand \bibnamefont  [1]{#1}%
\providecommand \bibfnamefont [1]{#1}%
\providecommand \citenamefont [1]{#1}%
\providecommand \href@noop [0]{\@secondoftwo}%
\providecommand \href [0]{\begingroup \@sanitize@url \@href}%
\providecommand \@href[1]{\@@startlink{#1}\@@href}%
\providecommand \@@href[1]{\endgroup#1\@@endlink}%
\providecommand \@sanitize@url [0]{\catcode `\\12\catcode `\$12\catcode `\&12\catcode `\#12\catcode `\^12\catcode `\_12\catcode `\%12\relax}%
\providecommand \@@startlink[1]{}%
\providecommand \@@endlink[0]{}%
\providecommand \url  [0]{\begingroup\@sanitize@url \@url }%
\providecommand \@url [1]{\endgroup\@href {#1}{\urlprefix }}%
\providecommand \urlprefix  [0]{URL }%
\providecommand \Eprint [0]{\href }%
\providecommand \doibase [0]{https://doi.org/}%
\providecommand \selectlanguage [0]{\@gobble}%
\providecommand \bibinfo  [0]{\@secondoftwo}%
\providecommand \bibfield  [0]{\@secondoftwo}%
\providecommand \translation [1]{[#1]}%
\providecommand \BibitemOpen [0]{}%
\providecommand \bibitemStop [0]{}%
\providecommand \bibitemNoStop [0]{.\EOS\space}%
\providecommand \EOS [0]{\spacefactor3000\relax}%
\providecommand \BibitemShut  [1]{\csname bibitem#1\endcsname}%
\let\auto@bib@innerbib\@empty
\bibitem [{\citenamefont {Feynman}\ \emph {et~al.}(2015)\citenamefont {Feynman}, \citenamefont {Leighton},\ and\ \citenamefont {Sands}}]{feynmanV3}%
  \BibitemOpen
  \bibfield  {author} {\bibinfo {author} {\bibfnamefont {R.~P.}\ \bibnamefont {Feynman}}, \bibinfo {author} {\bibfnamefont {R.~B.}\ \bibnamefont {Leighton}},\ and\ \bibinfo {author} {\bibfnamefont {M.}~\bibnamefont {Sands}},\ }\href@noop {} {\emph {\bibinfo {title} {The Feynman lectures on physics, Vol. III: Quantum Mechanics}}},\ Vol.~\bibinfo {volume} {3}\ (\bibinfo  {publisher} {Basic books},\ \bibinfo {year} {2015})\BibitemShut {NoStop}%
\bibitem [{\citenamefont {Richards}(2012)}]{richards2012budget}%
  \BibitemOpen
  \bibfield  {author} {\bibinfo {author} {\bibfnamefont {W.~J.}\ \bibnamefont {Richards}},\ }\bibfield  {title} {\bibinfo {title} {Budget cuts leave us science lagging},\ }\href@noop {} {\bibfield  {journal} {\bibinfo  {journal} {Nature}\ }\textbf {\bibinfo {volume} {488}},\ \bibinfo {pages} {32} (\bibinfo {year} {2012})}\BibitemShut {NoStop}%
\bibitem [{\citenamefont {Odom}\ \emph {et~al.}(2006)\citenamefont {Odom}, \citenamefont {Hanneke}, \citenamefont {D'Urso},\ and\ \citenamefont {Gabrielse}}]{electrong}%
  \BibitemOpen
  \bibfield  {author} {\bibinfo {author} {\bibfnamefont {B.}~\bibnamefont {Odom}}, \bibinfo {author} {\bibfnamefont {D.}~\bibnamefont {Hanneke}}, \bibinfo {author} {\bibfnamefont {B.}~\bibnamefont {D'Urso}},\ and\ \bibinfo {author} {\bibfnamefont {G.}~\bibnamefont {Gabrielse}},\ }\bibfield  {title} {\bibinfo {title} {New measurement of the electron magnetic moment using a one-electron quantum cyclotron},\ }\href {https://doi.org/10.1103/PhysRevLett.97.030801} {\bibfield  {journal} {\bibinfo  {journal} {Phys. Rev. Lett.}\ }\textbf {\bibinfo {volume} {97}},\ \bibinfo {pages} {030801} (\bibinfo {year} {2006})}\BibitemShut {NoStop}%
\bibitem [{\citenamefont {Deleglise}\ \emph {et~al.}(2008)\citenamefont {Deleglise}, \citenamefont {Dotsenko}, \citenamefont {Sayrin}, \citenamefont {Bernu}, \citenamefont {Brune}, \citenamefont {Raimond},\ and\ \citenamefont {Haroche}}]{deleglise2008}%
  \BibitemOpen
  \bibfield  {author} {\bibinfo {author} {\bibfnamefont {S.}~\bibnamefont {Deleglise}}, \bibinfo {author} {\bibfnamefont {I.}~\bibnamefont {Dotsenko}}, \bibinfo {author} {\bibfnamefont {C.}~\bibnamefont {Sayrin}}, \bibinfo {author} {\bibfnamefont {J.}~\bibnamefont {Bernu}}, \bibinfo {author} {\bibfnamefont {M.}~\bibnamefont {Brune}}, \bibinfo {author} {\bibfnamefont {J.-M.}\ \bibnamefont {Raimond}},\ and\ \bibinfo {author} {\bibfnamefont {S.}~\bibnamefont {Haroche}},\ }\bibfield  {title} {\bibinfo {title} {Reconstruction of non-classical cavity field states with snapshots of their decoherence},\ }\href@noop {} {\bibfield  {journal} {\bibinfo  {journal} {Nature}\ }\textbf {\bibinfo {volume} {455}},\ \bibinfo {pages} {510} (\bibinfo {year} {2008})}\BibitemShut {NoStop}%
\bibitem [{\citenamefont {Eibenberger}\ \emph {et~al.}(2013)\citenamefont {Eibenberger}, \citenamefont {Gerlich}, \citenamefont {Arndt}, \citenamefont {Mayor},\ and\ \citenamefont {T{\"u}xen}}]{Eibenberger}%
  \BibitemOpen
  \bibfield  {author} {\bibinfo {author} {\bibfnamefont {S.}~\bibnamefont {Eibenberger}}, \bibinfo {author} {\bibfnamefont {S.}~\bibnamefont {Gerlich}}, \bibinfo {author} {\bibfnamefont {M.}~\bibnamefont {Arndt}}, \bibinfo {author} {\bibfnamefont {M.}~\bibnamefont {Mayor}},\ and\ \bibinfo {author} {\bibfnamefont {J.}~\bibnamefont {T{\"u}xen}},\ }\bibfield  {title} {\bibinfo {title} {Matter--wave interference of particles selected from a molecular library with masses exceeding 10000 amu},\ }\href@noop {} {\bibfield  {journal} {\bibinfo  {journal} {Physical Chemistry Chemical Physics}\ }\textbf {\bibinfo {volume} {15}},\ \bibinfo {pages} {14696} (\bibinfo {year} {2013})}\BibitemShut {NoStop}%
\bibitem [{\citenamefont {Yu}\ \emph {et~al.}(2020)\citenamefont {Yu}, \citenamefont {McCuller}, \citenamefont {Tse}, \citenamefont {Kijbunchoo}, \citenamefont {Barsotti},\ and\ \citenamefont {Mavalvala}}]{haochun22}%
  \BibitemOpen
  \bibfield  {author} {\bibinfo {author} {\bibfnamefont {H.}~\bibnamefont {Yu}}, \bibinfo {author} {\bibfnamefont {L.}~\bibnamefont {McCuller}}, \bibinfo {author} {\bibfnamefont {M.}~\bibnamefont {Tse}}, \bibinfo {author} {\bibfnamefont {N.}~\bibnamefont {Kijbunchoo}}, \bibinfo {author} {\bibfnamefont {L.}~\bibnamefont {Barsotti}},\ and\ \bibinfo {author} {\bibfnamefont {N.}~\bibnamefont {Mavalvala}},\ }\bibfield  {title} {\bibinfo {title} {{Quantum correlations between light and the kilogram-mass mirrors of LIGO}},\ }\href@noop {} {\bibfield  {journal} {\bibinfo  {journal} {Nature}\ }\textbf {\bibinfo {volume} {583}},\ \bibinfo {pages} {43} (\bibinfo {year} {2020})}\BibitemShut {NoStop}%
\bibitem [{\citenamefont {Arute}\ \emph {et~al.}(2019)\citenamefont {Arute}, \citenamefont {Arya}, \citenamefont {Babbush}, \citenamefont {Bacon}, \citenamefont {Bardin}, \citenamefont {Barends}, \citenamefont {Biswas}, \citenamefont {Boixo}, \citenamefont {Brandao}, \citenamefont {Buell} \emph {et~al.}}]{google}%
  \BibitemOpen
  \bibfield  {author} {\bibinfo {author} {\bibfnamefont {F.}~\bibnamefont {Arute}}, \bibinfo {author} {\bibfnamefont {K.}~\bibnamefont {Arya}}, \bibinfo {author} {\bibfnamefont {R.}~\bibnamefont {Babbush}}, \bibinfo {author} {\bibfnamefont {D.}~\bibnamefont {Bacon}}, \bibinfo {author} {\bibfnamefont {J.~C.}\ \bibnamefont {Bardin}}, \bibinfo {author} {\bibfnamefont {R.}~\bibnamefont {Barends}}, \bibinfo {author} {\bibfnamefont {R.}~\bibnamefont {Biswas}}, \bibinfo {author} {\bibfnamefont {S.}~\bibnamefont {Boixo}}, \bibinfo {author} {\bibfnamefont {F.~G.}\ \bibnamefont {Brandao}}, \bibinfo {author} {\bibfnamefont {D.~A.}\ \bibnamefont {Buell}}, \emph {et~al.},\ }\bibfield  {title} {\bibinfo {title} {Quantum supremacy using a programmable superconducting processor},\ }\href@noop {} {\bibfield  {journal} {\bibinfo  {journal} {Nature}\ }\textbf {\bibinfo {volume} {574}},\ \bibinfo {pages} {505} (\bibinfo {year} {2019})}\BibitemShut {NoStop}%
\bibitem [{\citenamefont {Brukner}(2017)}]{brukner2017}%
  \BibitemOpen
  \bibfield  {author} {\bibinfo {author} {\bibfnamefont {{\v{C}}.}~\bibnamefont {Brukner}},\ }\bibfield  {title} {\bibinfo {title} {On the quantum measurement problem},\ }\href@noop {} {\bibfield  {journal} {\bibinfo  {journal} {Quantum [un] speakables II: half a century of Bell's theorem}\ ,\ \bibinfo {pages} {95}} (\bibinfo {year} {2017})}\BibitemShut {NoStop}%
\bibitem [{\citenamefont {Wigner}(1983)}]{Wigner1967}%
  \BibitemOpen
  \bibfield  {author} {\bibinfo {author} {\bibfnamefont {E.~P.}\ \bibnamefont {Wigner}},\ }\bibfield  {title} {\bibinfo {title} {Remarks on the mind-body question},\ }in\ \href@noop {} {\emph {\bibinfo {booktitle} {Quantum Theory and Measurement}}},\ \bibinfo {editor} {edited by\ \bibinfo {editor} {\bibfnamefont {J.~A.}\ \bibnamefont {Wheeler}}\ and\ \bibinfo {editor} {\bibfnamefont {W.~H.}\ \bibnamefont {Zurek}}}\ (\bibinfo  {publisher} {Princeton University Press},\ \bibinfo {address} {Princeton, NJ},\ \bibinfo {year} {1983})\ pp.\ \bibinfo {pages} {168--181},\ \bibinfo {note} {originally published in \emph{The Scientist Speculates}, edited by I. J. Good, London: Heinemann, 1961, pp. 284--302}\BibitemShut {NoStop}%
\bibitem [{\citenamefont {von Neumann}(1955)}]{vonNeumann1955}%
  \BibitemOpen
  \bibfield  {author} {\bibinfo {author} {\bibfnamefont {J.}~\bibnamefont {von Neumann}},\ }\href@noop {} {\emph {\bibinfo {title} {Mathematical Foundations of Quantum Mechanics}}}\ (\bibinfo  {publisher} {Princeton University Press},\ \bibinfo {address} {Princeton, NJ},\ \bibinfo {year} {1955})\BibitemShut {NoStop}%
\bibitem [{\citenamefont {Schmid}\ \emph {et~al.}(2023)\citenamefont {Schmid}, \citenamefont {Y{\=\i}ng},\ and\ \citenamefont {Leifer}}]{schmid2023review}%
  \BibitemOpen
  \bibfield  {author} {\bibinfo {author} {\bibfnamefont {D.}~\bibnamefont {Schmid}}, \bibinfo {author} {\bibfnamefont {Y.}~\bibnamefont {Y{\=\i}ng}},\ and\ \bibinfo {author} {\bibfnamefont {M.}~\bibnamefont {Leifer}},\ }\bibfield  {title} {\bibinfo {title} {A review and analysis of six extended wigner's friend arguments},\ }\href@noop {} {\bibfield  {journal} {\bibinfo  {journal} {arXiv preprint arXiv:2308.16220}\ } (\bibinfo {year} {2023})}\BibitemShut {NoStop}%
\bibitem [{\citenamefont {Deutsch}(1985)}]{Deutsch1985}%
  \BibitemOpen
  \bibfield  {author} {\bibinfo {author} {\bibfnamefont {D.}~\bibnamefont {Deutsch}},\ }\bibfield  {title} {\bibinfo {title} {Quantum theory as a universal physical theory},\ }\href {https://doi.org/10.1007/bf00670071} {\bibfield  {journal} {\bibinfo  {journal} {International Journal of Theoretical Physics}\ }\textbf {\bibinfo {volume} {24}},\ \bibinfo {pages} {1–41} (\bibinfo {year} {1985})}\BibitemShut {NoStop}%
\bibitem [{\citenamefont {Bennett}(1982)}]{bennett1982thermodynamics}%
  \BibitemOpen
  \bibfield  {author} {\bibinfo {author} {\bibfnamefont {C.~H.}\ \bibnamefont {Bennett}},\ }\bibfield  {title} {\bibinfo {title} {The thermodynamics of computation—a review},\ }\href@noop {} {\bibfield  {journal} {\bibinfo  {journal} {International Journal of Theoretical Physics}\ }\textbf {\bibinfo {volume} {21}},\ \bibinfo {pages} {905} (\bibinfo {year} {1982})}\BibitemShut {NoStop}%
\bibitem [{\citenamefont {Schwarzhans}\ \emph {et~al.}(2023)\citenamefont {Schwarzhans}, \citenamefont {Binder}, \citenamefont {Huber},\ and\ \citenamefont {Lock}}]{viennaGuys}%
  \BibitemOpen
  \bibfield  {author} {\bibinfo {author} {\bibfnamefont {E.}~\bibnamefont {Schwarzhans}}, \bibinfo {author} {\bibfnamefont {F.~C.}\ \bibnamefont {Binder}}, \bibinfo {author} {\bibfnamefont {M.}~\bibnamefont {Huber}},\ and\ \bibinfo {author} {\bibfnamefont {M.~P.~E.}\ \bibnamefont {Lock}},\ }\href {https://arxiv.org/abs/2302.11253} {\bibinfo {title} {Quantum measurements and equilibration: the emergence of objective reality via entropy maximisation}} (\bibinfo {year} {2023}),\ \Eprint {https://arxiv.org/abs/2302.11253} {arXiv:2302.11253 [quant-ph]} \BibitemShut {NoStop}%
\bibitem [{\citenamefont {Engineer}\ \emph {et~al.}(2024)\citenamefont {Engineer}, \citenamefont {Rivlin}, \citenamefont {Wollmann}, \citenamefont {Malik},\ and\ \citenamefont {Lock}}]{engineer2024}%
  \BibitemOpen
  \bibfield  {author} {\bibinfo {author} {\bibfnamefont {S.}~\bibnamefont {Engineer}}, \bibinfo {author} {\bibfnamefont {T.}~\bibnamefont {Rivlin}}, \bibinfo {author} {\bibfnamefont {S.}~\bibnamefont {Wollmann}}, \bibinfo {author} {\bibfnamefont {M.}~\bibnamefont {Malik}},\ and\ \bibinfo {author} {\bibfnamefont {M.~P.~E.}\ \bibnamefont {Lock}},\ }\href {https://arxiv.org/abs/2403.18016} {\bibinfo {title} {Equilibration of objective observables in a dynamical model of quantum measurements}} (\bibinfo {year} {2024}),\ \Eprint {https://arxiv.org/abs/2403.18016} {arXiv:2403.18016 [quant-ph]} \BibitemShut {NoStop}%
\bibitem [{\citenamefont {Lebowitz}(2007)}]{lebowitz2007time}%
  \BibitemOpen
  \bibfield  {author} {\bibinfo {author} {\bibfnamefont {J.~L.}\ \bibnamefont {Lebowitz}},\ }\bibinfo {title} {Boltzmann’s legacy}\ (\bibinfo  {publisher} {European Mathematical Society Zurich},\ \bibinfo {year} {2007})\ Chap.\ \bibinfo {chapter} {From time-symmetric microscopic dynamics to time-asymmetric macroscopic behavior: An overview}, pp.\ \bibinfo {pages} {63--88}\BibitemShut {NoStop}%
\bibitem [{\citenamefont {Brown}\ \emph {et~al.}(2009)\citenamefont {Brown}, \citenamefont {Myrvold},\ and\ \citenamefont {Uffink}}]{brown09}%
  \BibitemOpen
  \bibfield  {author} {\bibinfo {author} {\bibfnamefont {H.~R.}\ \bibnamefont {Brown}}, \bibinfo {author} {\bibfnamefont {W.}~\bibnamefont {Myrvold}},\ and\ \bibinfo {author} {\bibfnamefont {J.}~\bibnamefont {Uffink}},\ }\bibfield  {title} {\bibinfo {title} {{Boltzmann's H-theorem, its discontents, and the birth of statistical mechanics}},\ }\href {https://doi.org/https://doi.org/10.1016/j.shpsb.2009.03.003} {\bibfield  {journal} {\bibinfo  {journal} {Studies in History and Philosophy of Science Part B: Studies in History and Philosophy of Modern Physics}\ }\textbf {\bibinfo {volume} {40}},\ \bibinfo {pages} {174} (\bibinfo {year} {2009})}\BibitemShut {NoStop}%
\bibitem [{\citenamefont {Breuer}\ and\ \citenamefont {Petruccione}(2002)}]{breuer2002}%
  \BibitemOpen
  \bibfield  {author} {\bibinfo {author} {\bibfnamefont {H.-P.}\ \bibnamefont {Breuer}}\ and\ \bibinfo {author} {\bibfnamefont {F.}~\bibnamefont {Petruccione}},\ }\href@noop {} {\emph {\bibinfo {title} {The theory of open quantum systems}}}\ (\bibinfo  {publisher} {Oxford University Press},\ \bibinfo {year} {2002})\BibitemShut {NoStop}%
\bibitem [{\citenamefont {Zurek}(2003)}]{zurekRMP}%
  \BibitemOpen
  \bibfield  {author} {\bibinfo {author} {\bibfnamefont {W.~H.}\ \bibnamefont {Zurek}},\ }\bibfield  {title} {\bibinfo {title} {Decoherence, einselection, and the quantum origins of the classical},\ }\href {https://doi.org/10.1103/RevModPhys.75.715} {\bibfield  {journal} {\bibinfo  {journal} {Rev. Mod. Phys.}\ }\textbf {\bibinfo {volume} {75}},\ \bibinfo {pages} {715} (\bibinfo {year} {2003})}\BibitemShut {NoStop}%
\bibitem [{\citenamefont {Maudlin}(1995)}]{maudlin1995three}%
  \BibitemOpen
  \bibfield  {author} {\bibinfo {author} {\bibfnamefont {T.}~\bibnamefont {Maudlin}},\ }\bibfield  {title} {\bibinfo {title} {Three measurement problems},\ }\href@noop {} {\bibfield  {journal} {\bibinfo  {journal} {Topoi}\ }\textbf {\bibinfo {volume} {14}},\ \bibinfo {pages} {7} (\bibinfo {year} {1995})}\BibitemShut {NoStop}%
\bibitem [{\citenamefont {Schlosshauer}(2005)}]{decoherence_RMP}%
  \BibitemOpen
  \bibfield  {author} {\bibinfo {author} {\bibfnamefont {M.}~\bibnamefont {Schlosshauer}},\ }\bibfield  {title} {\bibinfo {title} {Decoherence, the measurement problem, and interpretations of quantum mechanics},\ }\href {https://doi.org/10.1103/RevModPhys.76.1267} {\bibfield  {journal} {\bibinfo  {journal} {Rev. Mod. Phys.}\ }\textbf {\bibinfo {volume} {76}},\ \bibinfo {pages} {1267} (\bibinfo {year} {2005})}\BibitemShut {NoStop}%
\bibitem [{\citenamefont {Schlosshauer}(2007)}]{schlosshauer}%
  \BibitemOpen
  \bibfield  {author} {\bibinfo {author} {\bibfnamefont {M.~A.}\ \bibnamefont {Schlosshauer}},\ }\href@noop {} {\emph {\bibinfo {title} {Decoherence: and the quantum-to-classical transition}}}\ (\bibinfo  {publisher} {Springer Science \& Business Media},\ \bibinfo {year} {2007})\BibitemShut {NoStop}%
\bibitem [{\citenamefont {Allahverdyan}\ \emph {et~al.}(2013)\citenamefont {Allahverdyan}, \citenamefont {Balian},\ and\ \citenamefont {Nieuwenhuizen}}]{allahverdyan2013}%
  \BibitemOpen
  \bibfield  {author} {\bibinfo {author} {\bibfnamefont {A.~E.}\ \bibnamefont {Allahverdyan}}, \bibinfo {author} {\bibfnamefont {R.}~\bibnamefont {Balian}},\ and\ \bibinfo {author} {\bibfnamefont {T.~M.}\ \bibnamefont {Nieuwenhuizen}},\ }\bibfield  {title} {\bibinfo {title} {Understanding quantum measurement from the solution of dynamical models},\ }\href@noop {} {\bibfield  {journal} {\bibinfo  {journal} {Physics Reports}\ }\textbf {\bibinfo {volume} {525}},\ \bibinfo {pages} {1} (\bibinfo {year} {2013})}\BibitemShut {NoStop}%
\bibitem [{\citenamefont {Carvalho}\ and\ \citenamefont {Correia}(2020)}]{gabriel2020}%
  \BibitemOpen
  \bibfield  {author} {\bibinfo {author} {\bibfnamefont {G.~D.}\ \bibnamefont {Carvalho}}\ and\ \bibinfo {author} {\bibfnamefont {P.~S.}\ \bibnamefont {Correia}},\ }\bibfield  {title} {\bibinfo {title} {Decay of quantumness in a measurement process: Action of a coarse-graining channel},\ }\href {https://doi.org/10.1103/PhysRevA.102.032217} {\bibfield  {journal} {\bibinfo  {journal} {Phys. Rev. A}\ }\textbf {\bibinfo {volume} {102}},\ \bibinfo {pages} {032217} (\bibinfo {year} {2020})}\BibitemShut {NoStop}%
\bibitem [{\citenamefont {Pranzini}\ and\ \citenamefont {Verrucchi}(2024)}]{pranzini2024}%
  \BibitemOpen
  \bibfield  {author} {\bibinfo {author} {\bibfnamefont {N.}~\bibnamefont {Pranzini}}\ and\ \bibinfo {author} {\bibfnamefont {P.}~\bibnamefont {Verrucchi}},\ }\bibfield  {title} {\bibinfo {title} {Premeasurement reliability and accessibility of quantum measurement apparatuses},\ }\href {https://doi.org/10.1103/PhysRevA.109.032203} {\bibfield  {journal} {\bibinfo  {journal} {Phys. Rev. A}\ }\textbf {\bibinfo {volume} {109}},\ \bibinfo {pages} {032203} (\bibinfo {year} {2024})}\BibitemShut {NoStop}%
\bibitem [{Note1()}]{Note1}%
  \BibitemOpen
  \bibinfo {note} {To be able to experimentally prepare the apparatus' initial state we must require $d_0/d_1\rightarrow 0$ as $\protect \textrm {poly}(1/d)$.}\BibitemShut {Stop}%
\bibitem [{\citenamefont {Del~Santo}\ and\ \citenamefont {Gisin}(2019)}]{santo2019}%
  \BibitemOpen
  \bibfield  {author} {\bibinfo {author} {\bibfnamefont {F.}~\bibnamefont {Del~Santo}}\ and\ \bibinfo {author} {\bibfnamefont {N.}~\bibnamefont {Gisin}},\ }\bibfield  {title} {\bibinfo {title} {Physics without determinism: Alternative interpretations of classical physics},\ }\href {https://doi.org/10.1103/PhysRevA.100.062107} {\bibfield  {journal} {\bibinfo  {journal} {Phys. Rev. A}\ }\textbf {\bibinfo {volume} {100}},\ \bibinfo {pages} {062107} (\bibinfo {year} {2019})}\BibitemShut {NoStop}%
\bibitem [{\citenamefont {Gisin}(2020)}]{gisin2020}%
  \BibitemOpen
  \bibfield  {author} {\bibinfo {author} {\bibfnamefont {N.}~\bibnamefont {Gisin}},\ }\bibfield  {title} {\bibinfo {title} {Mathematical languages shape our understanding of time in physics},\ }\href {https://doi.org/10.1038/s41567-019-0748-5} {\bibfield  {journal} {\bibinfo  {journal} {Nature Physics}\ }\textbf {\bibinfo {volume} {16}},\ \bibinfo {pages} {114} (\bibinfo {year} {2020})}\BibitemShut {NoStop}%
\bibitem [{\citenamefont {Tasaki}(1998)}]{tasaki98}%
  \BibitemOpen
  \bibfield  {author} {\bibinfo {author} {\bibfnamefont {H.}~\bibnamefont {Tasaki}},\ }\bibfield  {title} {\bibinfo {title} {From quantum dynamics to the canonical distribution: General picture and a rigorous example},\ }\href {https://doi.org/10.1103/PhysRevLett.80.1373} {\bibfield  {journal} {\bibinfo  {journal} {Phys. Rev. Lett.}\ }\textbf {\bibinfo {volume} {80}},\ \bibinfo {pages} {1373} (\bibinfo {year} {1998})}\BibitemShut {NoStop}%
\bibitem [{\citenamefont {Reimann}(2008)}]{reimann2008}%
  \BibitemOpen
  \bibfield  {author} {\bibinfo {author} {\bibfnamefont {P.}~\bibnamefont {Reimann}},\ }\bibfield  {title} {\bibinfo {title} {Foundation of statistical mechanics under experimentally realistic conditions},\ }\href {https://doi.org/10.1103/PhysRevLett.101.190403} {\bibfield  {journal} {\bibinfo  {journal} {Phys. Rev. Lett.}\ }\textbf {\bibinfo {volume} {101}},\ \bibinfo {pages} {190403} (\bibinfo {year} {2008})}\BibitemShut {NoStop}%
\bibitem [{\citenamefont {Linden}\ \emph {et~al.}(2009)\citenamefont {Linden}, \citenamefont {Popescu}, \citenamefont {Short},\ and\ \citenamefont {Winter}}]{linden09}%
  \BibitemOpen
  \bibfield  {author} {\bibinfo {author} {\bibfnamefont {N.}~\bibnamefont {Linden}}, \bibinfo {author} {\bibfnamefont {S.}~\bibnamefont {Popescu}}, \bibinfo {author} {\bibfnamefont {A.~J.}\ \bibnamefont {Short}},\ and\ \bibinfo {author} {\bibfnamefont {A.}~\bibnamefont {Winter}},\ }\bibfield  {title} {\bibinfo {title} {Quantum mechanical evolution towards thermal equilibrium},\ }\href {https://doi.org/10.1103/PhysRevE.79.061103} {\bibfield  {journal} {\bibinfo  {journal} {Phys. Rev. E}\ }\textbf {\bibinfo {volume} {79}},\ \bibinfo {pages} {061103} (\bibinfo {year} {2009})}\BibitemShut {NoStop}%
\bibitem [{\citenamefont {Gogolin}\ and\ \citenamefont {Eisert}(2016)}]{gogolin2016}%
  \BibitemOpen
  \bibfield  {author} {\bibinfo {author} {\bibfnamefont {C.}~\bibnamefont {Gogolin}}\ and\ \bibinfo {author} {\bibfnamefont {J.}~\bibnamefont {Eisert}},\ }\bibfield  {title} {\bibinfo {title} {Equilibration, thermalisation, and the emergence of statistical mechanics in closed quantum systems},\ }\href@noop {} {\bibfield  {journal} {\bibinfo  {journal} {Reports on Progress in Physics}\ }\textbf {\bibinfo {volume} {79}},\ \bibinfo {pages} {056001} (\bibinfo {year} {2016})}\BibitemShut {NoStop}%
\bibitem [{\citenamefont {Short}(2011)}]{Short2011}%
  \BibitemOpen
  \bibfield  {author} {\bibinfo {author} {\bibfnamefont {A.~J.}\ \bibnamefont {Short}},\ }\bibfield  {title} {\bibinfo {title} {Equilibration of quantum systems and subsystems},\ }\href {https://doi.org/10.1088/1367-2630/13/5/053009} {\bibfield  {journal} {\bibinfo  {journal} {New Journal of Physics}\ }\textbf {\bibinfo {volume} {13}},\ \bibinfo {pages} {053009} (\bibinfo {year} {2011})}\BibitemShut {NoStop}%
\bibitem [{\citenamefont {Short}\ and\ \citenamefont {Farrelly}(2012)}]{Short2012}%
  \BibitemOpen
  \bibfield  {author} {\bibinfo {author} {\bibfnamefont {A.~J.}\ \bibnamefont {Short}}\ and\ \bibinfo {author} {\bibfnamefont {T.~C.}\ \bibnamefont {Farrelly}},\ }\bibfield  {title} {\bibinfo {title} {Quantum equilibration in finite time},\ }\href {https://doi.org/10.1088/1367-2630/14/1/013063} {\bibfield  {journal} {\bibinfo  {journal} {New Journal of Physics}\ }\textbf {\bibinfo {volume} {14}},\ \bibinfo {pages} {013063} (\bibinfo {year} {2012})}\BibitemShut {NoStop}%
\bibitem [{\citenamefont {Forrester}(2010)}]{Forrester2010}%
  \BibitemOpen
  \bibfield  {author} {\bibinfo {author} {\bibfnamefont {P.~J.}\ \bibnamefont {Forrester}},\ }\href {https://doi.org/10.1515/9781400835416} {\emph {\bibinfo {title} {Log-Gases and Random Matrices (LMS-34)}}}\ (\bibinfo  {publisher} {Princeton University Press},\ \bibinfo {year} {2010})\BibitemShut {NoStop}%
\bibitem [{\citenamefont {Duarte}\ \emph {et~al.}(2017)\citenamefont {Duarte}, \citenamefont {Carvalho}, \citenamefont {Bernardes},\ and\ \citenamefont {de~Melo}}]{cris2017}%
  \BibitemOpen
  \bibfield  {author} {\bibinfo {author} {\bibfnamefont {C.}~\bibnamefont {Duarte}}, \bibinfo {author} {\bibfnamefont {G.~D.}\ \bibnamefont {Carvalho}}, \bibinfo {author} {\bibfnamefont {N.~K.}\ \bibnamefont {Bernardes}},\ and\ \bibinfo {author} {\bibfnamefont {F.}~\bibnamefont {de~Melo}},\ }\bibfield  {title} {\bibinfo {title} {Emerging dynamics arising from coarse-grained quantum systems},\ }\href {https://doi.org/10.1103/PhysRevA.96.032113} {\bibfield  {journal} {\bibinfo  {journal} {Phys. Rev. A}\ }\textbf {\bibinfo {volume} {96}},\ \bibinfo {pages} {032113} (\bibinfo {year} {2017})}\BibitemShut {NoStop}%
\bibitem [{\citenamefont {Vallejos}\ \emph {et~al.}(2022)\citenamefont {Vallejos}, \citenamefont {Correia}, \citenamefont {Obando}, \citenamefont {O'Neill}, \citenamefont {Tacla},\ and\ \citenamefont {de~Melo}}]{vallejos2022}%
  \BibitemOpen
  \bibfield  {author} {\bibinfo {author} {\bibfnamefont {R.~O.}\ \bibnamefont {Vallejos}}, \bibinfo {author} {\bibfnamefont {P.~S.}\ \bibnamefont {Correia}}, \bibinfo {author} {\bibfnamefont {P.~C.}\ \bibnamefont {Obando}}, \bibinfo {author} {\bibfnamefont {N.~M.}\ \bibnamefont {O'Neill}}, \bibinfo {author} {\bibfnamefont {A.~B.}\ \bibnamefont {Tacla}},\ and\ \bibinfo {author} {\bibfnamefont {F.}~\bibnamefont {de~Melo}},\ }\bibfield  {title} {\bibinfo {title} {Quantum state inference from coarse-grained descriptions: Analysis and an application to quantum thermodynamics},\ }\href {https://doi.org/10.1103/PhysRevA.106.012219} {\bibfield  {journal} {\bibinfo  {journal} {Phys. Rev. A}\ }\textbf {\bibinfo {volume} {106}},\ \bibinfo {pages} {012219} (\bibinfo {year} {2022})}\BibitemShut {NoStop}%
\bibitem [{\citenamefont {Correia}\ \emph {et~al.}(2024)\citenamefont {Correia}, \citenamefont {Carvalho}, \citenamefont {de~Oliveira}, \citenamefont {Vallejos},\ and\ \citenamefont {de~Melo}}]{pedrinho_PRL24}%
  \BibitemOpen
  \bibfield  {author} {\bibinfo {author} {\bibfnamefont {P.~S.}\ \bibnamefont {Correia}}, \bibinfo {author} {\bibfnamefont {G.~D.}\ \bibnamefont {Carvalho}}, \bibinfo {author} {\bibfnamefont {T.~R.}\ \bibnamefont {de~Oliveira}}, \bibinfo {author} {\bibfnamefont {R.~O.}\ \bibnamefont {Vallejos}},\ and\ \bibinfo {author} {\bibfnamefont {F.}~\bibnamefont {de~Melo}},\ }\bibfield  {title} {\bibinfo {title} {Canonical typicality under general quantum channels},\ }\href {https://doi.org/10.1103/PhysRevLett.133.060401} {\bibfield  {journal} {\bibinfo  {journal} {Phys. Rev. Lett.}\ }\textbf {\bibinfo {volume} {133}},\ \bibinfo {pages} {060401} (\bibinfo {year} {2024})}\BibitemShut {NoStop}%
\bibitem [{SM(2025)}]{SM}%
  \BibitemOpen
  \href@noop {} {\bibinfo {title} {Supplemental material}},\ \bibinfo {howpublished} {Available at \url{http://example.com/supplemental-material}} (\bibinfo {year} {2025})\BibitemShut {NoStop}%
\bibitem [{\citenamefont {Mele}(2024)}]{Mele2024}%
  \BibitemOpen
  \bibfield  {author} {\bibinfo {author} {\bibfnamefont {A.~A.}\ \bibnamefont {Mele}},\ }\bibfield  {title} {\bibinfo {title} {{Introduction to Haar Measure Tools in Quantum Information: A Beginner's Tutorial}},\ }\href {https://doi.org/10.22331/q-2024-05-08-1340} {\bibfield  {journal} {\bibinfo  {journal} {Quantum}\ }\textbf {\bibinfo {volume} {8}},\ \bibinfo {pages} {1340} (\bibinfo {year} {2024})}\BibitemShut {NoStop}%
\bibitem [{\citenamefont {Fukuda}\ \emph {et~al.}(2019)\citenamefont {Fukuda}, \citenamefont {König},\ and\ \citenamefont {Nechita}}]{RTNI}%
  \BibitemOpen
  \bibfield  {author} {\bibinfo {author} {\bibfnamefont {M.}~\bibnamefont {Fukuda}}, \bibinfo {author} {\bibfnamefont {R.}~\bibnamefont {König}},\ and\ \bibinfo {author} {\bibfnamefont {I.}~\bibnamefont {Nechita}},\ }\bibfield  {title} {\bibinfo {title} {{RTNI—A symbolic integrator for Haar-random tensor networks}},\ }\href {https://doi.org/10.1088/1751-8121/ab434b} {\bibfield  {journal} {\bibinfo  {journal} {Journal of Physics A: Mathematical and Theoretical}\ }\textbf {\bibinfo {volume} {52}},\ \bibinfo {pages} {425303} (\bibinfo {year} {2019})}\BibitemShut {NoStop}%
\bibitem [{\citenamefont {Meckes}(2019)}]{Meckes2019}%
  \BibitemOpen
  \bibfield  {author} {\bibinfo {author} {\bibfnamefont {E.~S.}\ \bibnamefont {Meckes}},\ }\href@noop {} {\emph {\bibinfo {title} {The Random Matrix Theory of the Classical Compact Groups}}},\ Cambridge Tracts in Mathematics\ (\bibinfo  {publisher} {Cambridge University Press},\ \bibinfo {year} {2019})\BibitemShut {NoStop}%
\bibitem [{\citenamefont {Popescu}\ \emph {et~al.}(2006)\citenamefont {Popescu}, \citenamefont {Short},\ and\ \citenamefont {Winter}}]{popescu2006}%
  \BibitemOpen
  \bibfield  {author} {\bibinfo {author} {\bibfnamefont {S.}~\bibnamefont {Popescu}}, \bibinfo {author} {\bibfnamefont {A.~t.}\ \bibnamefont {Short}},\ and\ \bibinfo {author} {\bibfnamefont {A.}~\bibnamefont {Winter}},\ }\bibfield  {title} {\bibinfo {title} {Entanglement and the foundations of statistical mechanics},\ }\href {https://doi.org/10.1038/nphys444} {\bibfield  {journal} {\bibinfo  {journal} {Nature Phys}\ }\textbf {\bibinfo {volume} {2}},\ \bibinfo {pages} {754} (\bibinfo {year} {2006})}\BibitemShut {NoStop}%
\bibitem [{\citenamefont {Low}(2009)}]{low2009}%
  \BibitemOpen
  \bibfield  {author} {\bibinfo {author} {\bibfnamefont {R.~A.}\ \bibnamefont {Low}},\ }\bibfield  {title} {\bibinfo {title} {Large deviation bounds for $k$-designs},\ }\href {https://doi.org/10.1098/rspa.2009.0232} {\bibfield  {journal} {\bibinfo  {journal} {Proceedings of the Royal Society A: Mathematical, Physical and Engineering Sciences}\ }\textbf {\bibinfo {volume} {465}},\ \bibinfo {pages} {3289} (\bibinfo {year} {2009})}\BibitemShut {NoStop}%
\bibitem [{\citenamefont {Darrigol}(2021)}]{darrigol2021boltzmann}%
  \BibitemOpen
  \bibfield  {author} {\bibinfo {author} {\bibfnamefont {O.}~\bibnamefont {Darrigol}},\ }\bibfield  {title} {\bibinfo {title} {{Boltzmann’s reply to the Loschmidt paradox: a commented translation}},\ }\href@noop {} {\bibfield  {journal} {\bibinfo  {journal} {The European Physical Journal H}\ }\textbf {\bibinfo {volume} {46}},\ \bibinfo {pages} {29} (\bibinfo {year} {2021})}\BibitemShut {NoStop}%
\bibitem [{\citenamefont {Binder}(2023)}]{binder2023reversibility}%
  \BibitemOpen
  \bibfield  {author} {\bibinfo {author} {\bibfnamefont {P.}~\bibnamefont {Binder}},\ }\bibfield  {title} {\bibinfo {title} {The reversibility paradox: Role of the velocity reversal step},\ }\href {https://doi.org/10.1007/s10773-023-05458-x} {\bibfield  {journal} {\bibinfo  {journal} {International Journal of Theoretical Physics}\ }\textbf {\bibinfo {volume} {62}},\ \bibinfo {pages} {200} (\bibinfo {year} {2023})}\BibitemShut {NoStop}%
\bibitem [{\citenamefont {Reimann}(2010)}]{Reimann_2010}%
  \BibitemOpen
  \bibfield  {author} {\bibinfo {author} {\bibfnamefont {P.}~\bibnamefont {Reimann}},\ }\bibfield  {title} {\bibinfo {title} {Canonical thermalization},\ }\href {https://doi.org/10.1088/1367-2630/12/5/055027} {\bibfield  {journal} {\bibinfo  {journal} {New Journal of Physics}\ }\textbf {\bibinfo {volume} {12}},\ \bibinfo {pages} {055027} (\bibinfo {year} {2010})}\BibitemShut {NoStop}%
\bibitem [{\citenamefont {Bong}\ \emph {et~al.}(2020)\citenamefont {Bong}, \citenamefont {Utreras-Alarc{\'o}n}, \citenamefont {Ghafari}, \citenamefont {Liang}, \citenamefont {Tischler}, \citenamefont {Cavalcanti}, \citenamefont {Pryde},\ and\ \citenamefont {Wiseman}}]{bong2020}%
  \BibitemOpen
  \bibfield  {author} {\bibinfo {author} {\bibfnamefont {K.-W.}\ \bibnamefont {Bong}}, \bibinfo {author} {\bibfnamefont {A.}~\bibnamefont {Utreras-Alarc{\'o}n}}, \bibinfo {author} {\bibfnamefont {F.}~\bibnamefont {Ghafari}}, \bibinfo {author} {\bibfnamefont {Y.-C.}\ \bibnamefont {Liang}}, \bibinfo {author} {\bibfnamefont {N.}~\bibnamefont {Tischler}}, \bibinfo {author} {\bibfnamefont {E.~G.}\ \bibnamefont {Cavalcanti}}, \bibinfo {author} {\bibfnamefont {G.~J.}\ \bibnamefont {Pryde}},\ and\ \bibinfo {author} {\bibfnamefont {H.~M.}\ \bibnamefont {Wiseman}},\ }\bibfield  {title} {\bibinfo {title} {{A strong no-go theorem on the Wigner’s friend paradox}},\ }\href@noop {} {\bibfield  {journal} {\bibinfo  {journal} {Nature Physics}\ }\textbf {\bibinfo {volume} {16}},\ \bibinfo {pages} {1199} (\bibinfo {year} {2020})}\BibitemShut {NoStop}%
\bibitem [{\citenamefont {Santo}\ \emph {et~al.}(2024)\citenamefont {Santo}, \citenamefont {Manzano},\ and\ \citenamefont {Brukner}}]{delsanto2024}%
  \BibitemOpen
  \bibfield  {author} {\bibinfo {author} {\bibfnamefont {F.~D.}\ \bibnamefont {Santo}}, \bibinfo {author} {\bibfnamefont {G.}~\bibnamefont {Manzano}},\ and\ \bibinfo {author} {\bibfnamefont {C.}~\bibnamefont {Brukner}},\ }\href {https://arxiv.org/abs/2407.06279} {\bibinfo {title} {{Wigner's friend scenarios: on what to condition and how to verify the predictions}}} (\bibinfo {year} {2024}),\ \Eprint {https://arxiv.org/abs/2407.06279} {arXiv:2407.06279 [quant-ph]} \BibitemShut {NoStop}%
\bibitem [{\citenamefont {Frauchiger}\ and\ \citenamefont {Renner}(2018)}]{frauchiger2018}%
  \BibitemOpen
  \bibfield  {author} {\bibinfo {author} {\bibfnamefont {D.}~\bibnamefont {Frauchiger}}\ and\ \bibinfo {author} {\bibfnamefont {R.}~\bibnamefont {Renner}},\ }\bibfield  {title} {\bibinfo {title} {Quantum theory cannot consistently describe the use of itself},\ }\href@noop {} {\bibfield  {journal} {\bibinfo  {journal} {Nature communications}\ }\textbf {\bibinfo {volume} {9}},\ \bibinfo {pages} {3711} (\bibinfo {year} {2018})}\BibitemShut {NoStop}%
\end{thebibliography}%

\clearpage


\section{Supplemental Material}

Here we give details of the calculations presented in the main-text. 
%
Before starting, notice that as $H_{[+1,0]}$ and $H_{[0,-1]}$ are independently taken from the GUE ensemble, besides being  individually non-degenerate and not having degenerate gaps, they also do not have equal eigenvalues nor eigenvectors. Let $S_{[+1,0]}=\{E_m^+\}_{m=1}^{d_0+d_1}$  be the spectrum of $H_{[+1,0]}$ and $\{\ket{E_m^+}\}_{m=1}^{d_0+d_1}$ its eigenvectors in $\mc{H}_{+1}\oplus \mc{H}_0$, and, similarly, $S_{[0,-1]}=\{E_n^-\}_{n=1}^{d_0+d_1}$ and  $\{\ket{E_n^-}\}_{n=1}^{d_0+d_1}\in \mc{H}_{0}\oplus \mc{H}_{-1}$ be the spectrum and eigenvectors of  $H_{[0,-1]}$. Then, the eigenvectors of $H$ with non-null eigenvalues  are $\{\ket{\phi^+_m}:=\ket{+}\otimes(\ket{E_m^+}\oplus 0)\}_{m=1}^{d_0+d_1}\cup\{\ket{\phi^-_n}:=\ket{-}\otimes(0\oplus\ket{E^-_n})\}_{n=1}^{d_0+d_1}$. 
Nevertheless, $H$ also has a degenerate subspace corresponding to the zero-energy eigenvalues, namely its kernel $\mc{K}(H)=\SpanOp\{ \ket{+}\otimes(0\oplus0\oplus \ket{\chi}),\; \ket{-}\otimes (\ket{\chi^\prime}\oplus 0\oplus 0) |\; \forall\; \ket{\chi}\in \mc{H}_{-1}, \forall\; \ket{\chi^\prime} \in \mc{H}_{+1}\}$. 
Given that the thermalization bounds~\eqref{eq:equilibration_bound} and \eqref{eq:distinguishability} require non-degeneracy, and that the zero eigeinvalue subspace does not play a role in the described dynamics, in all the following calculations we will be dealing with the Hamiltonian
\[
H_* = \sum_{m=1}^{d_0+d_1} E_m^+ \proj{\phi^+_m}+\sum_{n=1}^{d_0+d1} E^-_n \proj{\phi^-_n},
\]
which is equivalent to $H$, in \eqref{eq:hamiltonian}, without its kernel, and such that $d_*:=\dim{H_*}= 2(d_1+d_0+d_1)- 2 d_1= 2(d_0+d_1)$.

\subsection{Evaluation of the effective dimension}

In the main-text we employed two results from the field of equilibration of closed quantum systems. The first one concerns the equilibration of an observable property, and it is explicitly written in Eq.\eqref{eq:equilibration_bound}. The second one concerns the equilibration of the state describing the system given a finite set of POVMs. It is a corollary of the bound for observables, and, as described in \cite{Short2011}, it reads in general as follows:
\begin{equation}
    \<\mc{D}_\mc{M}(\rho_t^H,\rho_\infty^H)\>_\infty\le \frac{\mc{S}(\mc{M})}{4\sqrt{d^H_\text{eff}(\rho_0)}}.
    \label{eq:distinguishability}
\end{equation}
Both bounds depend on the effective dimension $d^H_\text{eff}(\rho_0)=1/(\sum_i \<E_i|\rho_0|E_i\>^2)$, with $\{\ket{E_i}\}$ the eigenvectors of $H$.

In order to evaluate the effective dimension for our model, we consider the Hamiltonian $H_*$ above, and the initial state $\rho_0=\psi\otimes (0\oplus\idty\oplus0)/d_0$. In this case:
\begin{equation}
    \frac{1}{d^{H_*}_\text{eff}(\rho_0)}=\sum_m \bra{\phi^+_m}\rho_0\ket{\phi^+_m}^2+ \sum_m \bra{\phi^-_m}\rho_0\ket{\phi^-_m}^2.
\end{equation}
Explicitly writing $\ket{\phi^+_m}=\ket{+}\otimes(\ket{u_m^+}\oplus \ket{z_m^+}\oplus0)$, and $\ket{\phi^-_m}=\ket{-}\otimes(0\oplus \ket{z_m^-}\oplus \ket{b_m^-})$, we get
\begin{equation}
    \frac{1}{d^{H_*}_\text{eff}(\rho_0)}=\frac{|c_+|^4}{d_0^2}\sum_{m=1}^{d_0+d_1} \bra{z^+_m} z^+_m\>^2+\frac{|c_-|^4}{d_0^2}\sum_{m=1}^{d_0+d_1} \bra{z^-_m} z^-_m\>^2.
\end{equation}
Note that the vectors $\ket{z_m^+}$ and $\ket{z_m^-}$ are sub-normalized.

For large apparatus' dimension, the eigenvectors of $H_{[+1,0]}$ and $H_{[0,-1]}$ are uniformly distributed. As such, fixing a basis for $\Cx^d$, the amplitude of each component of any eigenvector $\ket{E_m^+}\in \mc{H}_{[+1,0]}$ will have, on average, absolute value approximately equal to $1/\sqrt{d_0+d_1}$. The same is true for all eigenvectors $\ket{E_n^-}\in \mc{H}_{[0,-1]}$.  Within this assumption, we  have $\bra{z^+_m} z^+_m\>\approx \bra{z^-_m} z^-_m\> \approx d_0/(d_0+d_1)$.  Therefore, for large $d$:
\begin{equation}
    \frac{1}{d^{H_*}_\text{eff}(\rho_0)}\approx \frac{|c_+|^4+|c_-|^4}{(d_0+d_1)} \le \frac{1}{d_0+d_1}, 
\end{equation}
where for the last inequality we used that $|c_+|^4+|c_-|^4 \le 1 $. With the above estimation for the effective dimension we reach results \eqref{eq:eq_probs} and \eqref{eq:bound_dist}.

\subsection{Equilibrium state: fixed Hamiltonian}

For a given Hamiltonian $H$ and initial state $\rho_0$, the equilibrium state is given by the time-averaged state $\rho_\infty^H=\lim_{T\rightarrow \infty} \frac{1}{T} \int_0^Tdt e^{-\ii H t} \rho_0 e^{\ii H t}$. If the Hamiltonian has no degenerate levels, nor degenerate gaps, such state can be equivalently obtained by the pinching map~\cite{gogolin2016}, which simply corresponds to the diagonal elements of the initial state in the Hamiltonian basis. 

Therefore, for our model, the equilibrium state for a fixed choice of $H_*$ and $\rho_0$ is given by the corresponding pinching map:
\[
\rho_\infty^H=\sum_{m=1}^{d_0+d_1} \bra{\phi_m^+} \rho_0 \ket{\phi_m^+} + \sum_{m=1}^{d_0+d_1} \bra{\phi_m^-} \rho_0 \ket{\phi_m^-}.
\]
Writing  $P^\pm_m= \proj{E_m^\pm}$ for the rank-1 energy projector, and explicitly using $\rho_0 =\psi \otimes (0\oplus \idty\oplus 0)/d_0$, we get:
\begin{multline}
   \rho_\infty^H = |c_+|^2\proj{+}\otimes \sum_{m=1}^{d_0+d_1} \left[ P_m^+\left(0\oplus\frac{\idty}{d_0}\right)P_m^+\right]\oplus 0 +\\
   +|c_-|^2\proj{-}\otimes \sum_{m=1}^{d_0+d_1} 0\oplus \left[ P_m^-\left(\frac{\idty}{d_0}\oplus 0\right)P_m^-\right].
   \label{eq_app:average_state_fixedH}
\end{multline}
This expression will be employed below.

\subsection{Typical Equilibrium State and Equilibrium Probabilities}

Starting with state \eqref{eq_app:average_state_fixedH}, we must evaluate the average over Hamiltonians $H_{[+1,0]}$ and $H_{[0,-1]}$ sampled from the GUE ensemble of the state $\rho_\infty^H$, i.e., we must evaluate $\overline{\rho_\infty}$:
\begin{multline*}
   \overline{\rho_\infty} = |c_+|^2\proj{+}\otimes \sum_{m=1}^{d_0+d_1} \overline{\left[ P_m^+\left(0\oplus\frac{\idty}{d_0}\right)P_m^+\right]}\oplus 0 +\\
   +|c_-|^2\proj{-}\otimes \sum_{m=1}^{d_0+d_1} 0\oplus \overline{\left[ P_m^-\left(\frac{\idty}{d_0}\oplus 0\right)P_m^-\right]}.
\end{multline*}
The average over Hamiltonians can be recast as an average over unitary matrices by observing that the eigenvectors associated with different choices of \( H_{[+1,0]} \) are related by unitary transformations. Moreover, since the eigenvectors of \( H_{[+1,0]} \) in the GUE ensemble are distributed according to the Haar measure, so are the unitary matrices that relate them (the same holds for \( H_{[0,-1]} \)).
Thus, using  well-known Haar integrals~\cite{Forrester2010,Meckes2019, Mele2024} we get:
\begin{equation}
\begin{split}
 \overline{\rho_\infty} &= 
   |c_+|^2 \proj{+} \otimes \frac{1}{d_0+d_1+1} 
 \left( 0 \oplus \frac{\idty_0}{d_0} + \idty_{[+1,0]} \right) \oplus 0 \\
 & + |c_-|^2 \proj{-} \otimes \frac{1}{d_0+d_1+1} 
 0 \oplus \left( \frac{\idty_0}{d_0} \oplus 0 + \idty_{[0,-1]} \right). \end{split}
\end{equation}
Regrouping terms in the last equation allows us to recover Eq.~(\ref{eq:fully_avg_state}) from the main text.

\subsection{Concentration bound for the equilibrium probabilities: Chebyshev's inequality}

In order to show the concentration of the equilibrium probabilities around their average values we employed the well-known Chebyshev's inequality:
\begin{lemma}
    Let $X$ be a random variable with mean $\mu$ and non-zero variance $\sigma^2$. Then, for all $\delta>0$:
        \[\pr\left(|X-\mu|\ge \delta\right)\le \frac{\sigma^2}{\delta^2}.\]
\end{lemma}

The random variables of interest are then $\pr(i|H)_\infty$, for $i\in\{+1,0,-1\}$, where $H_{[+1,0]}$ and $H_{[0,-1]}$ are sampled uniformly from the GUE ensemble and form the total Hamiltonian $H$. 

To evaluate the average values $\overline{\pr(i|H)_\infty}$, given the average typical state, Eq.~(\ref{eq:fully_avg_state}), is immediate as $\overline{\pr(i|H)_\infty}=\tr(\idty\otimes \Pi_i \overline{\rho_\infty})$. This leads directly to the expressions in \eqref{eq:typical_probs}.

For the variances we need to evaluate the average of $\pr(i|H)_\infty^2$. For example, for the outcome $i=+1$ such expression reads:
\begin{equation*}
\begin{split}
    \pr(+1|H_{[+1,0]})_\infty^2=|c_+|^4\sum_{m,n=1}^{d_0+d_1} \bra{m^+} U^\dagger(\Pi_{+1}\oplus 0) U\ket{m^+} \\
    \times  \bra{m^+}U^\dagger  (0 \oplus \frac{\idty}{d_0})U\ket{m^+}\bra{n^+}U^\dagger (\Pi_{+1}\oplus 0) U \ket{n^+}\\
    \times \bra{n^+}U^\dagger(0 \oplus \frac{\idty}{d_0})U\ket{n^+},
\end{split}
\end{equation*}
where $\{\ket{m^+}\}$ is an orthonormal basis for $\mc{H}_{+1}\oplus \mc{H}_0$, and $U$ is the unitary transformation from such basis to the eigen-basis of a given $H_{[+1,0]}$. A similar expression holds for $\pr(-1|H_{[0,-1]})_\infty^2$.

The evaluation of the average of such expressions can be performed by standard techniques shown, for instance, in \cite{Meckes2019,Mele2024}. Nevertheless, due to the large number of terms, we employed the Random Tensor Network Integrator  symbolic library~\cite{RTNI}. With that we obtain:
\small
\begin{equation*}
   \sigma_{\pm 1}^2 =  \frac{2 |c_\pm|^4(d_0+1) d_1 (d_1+1)}{d_0 (d_0+d_1) (d_0+d_1+1)^2 (d_0+d_1+2) (d_0+d_1+3)}.
\end{equation*}
\normalsize
Given that $\sum_i \pr(i|H)_\infty=1$, we have that $\sigma_0^2\le 2(\sigma_{+1}^2+\sigma_{-1}^2)$, we get 
\small
\begin{equation*}
   \sigma_{0}^2 \le  \frac{4 (|c_{+1}|^4+|c_{-1}|^4)(d_0+1) d_1 (d_1+1)}{d_0 (d_0+d_1) (d_0+d_1+1)^2 (d_0+d_1+2) (d_0+d_1+3)}.
\end{equation*}
\normalsize
Finally, using that $|c_\pm|^4\le 1$ and $|c_+|^4+|c_-|^4\le 1$, we obtain a state-independent upper bound for the variances:
\small
\begin{equation*}
\begin{split}
\sigma_{i}^2 &\le  \frac{4 (d_0+1) d_1 (d_1+1)}{d_0 (d_0+d_1) (d_0+d_1+1)^2 (d_0+d_1+2) (d_0+d_1+3)},\\
    & =\frac{4  (d_0+1)}{d_0 d_1^3}+O\left(\frac{1}{d_1^4}\right),
\end{split}
\end{equation*}
\normalsize
which leads to the result reported in~\eqref{eq:chebyshev_probs}.

\subsection{Concentration bounds: Levy's Lemma}

The specific form of Levy's lemma we used can be found, for instance, in \cite{low2009} and reads:
\begin{lemma}
Let $f$ be a $\eta$-Lipschitz function on $D$-dimensional unitary matrices, with mean over the Haar measure $\overline{f}$. Then, $\forall \delta\ge 0$:
\begin{equation}
\pr_{U\sim \mu_{\text{Haar}}}(|f(U)-\overline{f}|\ge \delta)\le 4 e^{-\frac{c D \delta^2}{\eta^2}}, 
\label{eq_app:levys_lemma}
\end{equation}
 where $c$ is a positive constant, and the Lipschitz constant is defined as $\eta = \sup_{U,V} |f(U)-f(V)|/||U-V||_2$.
\end{lemma}

We want to apply Levy's lemma to the function  $||\rho_\infty^H- \overline{\rho_\infty}||_2^2$ when sampling $H_{[+1,0]}$ and $H_{[0,-1]}$ from the GUE ensemble. To write such functions as  requested in~\eqref{eq_app:levys_lemma}, we proceed in the same way as explained in the evaluation of the average typical state. It turns out that by exployting the expressions for $\rho_\infty^H$ and  $\overline{\rho_\infty}$, the only term  in $||\rho_\infty^H- \overline{\rho_\infty}||_2^2=\tr[(\rho_\infty^H)^2]- 2\tr(\rho_\infty^H\overline{\rho_\infty})+\tr[(\overline{\rho_\infty})^2]$ that depends on the choice of Hamiltonians is  $\tr[(\rho_\infty^H)^2]$, with the other terms constant. Explicitly evaluating the non-constant term, we can write $\tr[(\rho_\infty^H)^2]= f^+(U)+ f^-(U^\prime)$, where:
\begin{multline*}
    f^+(U)= |c_+|^4 \sum_{m=1}^{d_0+d_1}\left[\bra{E_m^+} U^\dagger \left(0\oplus \frac{\idty}{d_0}\right) U \ket{E_m^+}\right]^2,\\
    f^-(U^\prime)= |c_-|^4 \sum_{m=1}^{d_0+d_1}\left[\bra{E_m^-} {U^\prime}^\dagger \left(\frac{\idty}{d_0}\oplus 0\right) U^\prime \ket{E_m^-}\right]^2,
\end{multline*}
with $U$ the unitary connecting the eigenvectors of different choices of $H_{[+1,0]}$, and, similarly, $U^\prime$ connects the eigenvectors of $H_{[0,-1]}$. If we are able to apply Levy's lemma for each of these functions individually, by the union bound we get an exponential concentration around the mean for $||\rho_\infty^H- \overline{\rho_\infty}||_2^2$.

Therefore, the only thing we have to show is that the functions $f^\pm$ are $\eta$-Lipschitz, with appropriate Lipschitz constants.  We then consider the Lipschitz continuity of the  function $f(U)= \sum_{m=1}^{d_0+d_1}\left[\bra{m} U^\dagger A U \ket{m} \right]^2$, with $A$  a trace-one Hermitian operator, and $\{\ket{m}\}$ an orthonormal basis for $\Cx^{d_0+d_1}$. Then:
\onecolumngrid
\begin{align*}
    |f(U)-f(V)|&= \Bigg| \sum_{m=1}^{d_0+d_1}\left[\bra{m} U^\dagger A U \ket{m} \right]^2 -\left[\bra{m} V^\dagger A V \ket{m} \right]^2\Bigg|\le \sum_{m=1}^{d_0+d_1}\Big| \left[\bra{m} U^\dagger A U \ket{m} \right]^2 -\left[\bra{m} V^\dagger A V \ket{m} \right]^2\Big|;\\
    &\le \sum_{m=1}^{d_0+d_1}\Big|\bra{m} U^\dagger A U \ket{m}(\bra{m} U^\dagger A U \ket{m}-\bra{m} V^\dagger A V \ket{m})\Big| + \Big|\bra{m} V^\dagger A V \ket{m}(\bra{m} V^\dagger A V \ket{m}-\bra{m} U^\dagger A U \ket{m})\Big|;\\
    &\le \left[\left(\sum_{m=1}^{d_0+d_1} \Big|\bra{m} U^\dagger A U \ket{m}\Big|^2\right)^\frac{1}{2}+\left(\sum_{m=1}^{d_0+d_1} \Big|\bra{m} V^\dagger A V \ket{m}\Big|^2\right)^\frac{1}{2}\right]\left(\sum_{m=1}^{d_0+d_1}\Big|\bra{m} U^\dagger A U \ket{m}-\bra{m} V^\dagger A V \ket{m}\Big|^2\right)^\frac{1}{2};\\
    &\le 2||A||_2\left(\sum_{m=1}^{d_0+d_1}\Big|\bra{m} U^\dagger A U \ket{m}-\bra{m} V^\dagger A V \ket{m}\Big|^2\right)^\frac{1}{2};\\
    &\le 2||A||_2^2\left(\sum_{m=1}^{d_0+d_1}|| U\proj{m} U^\dagger - V \proj{m}V^\dagger||_2^2\right)^\frac{1}{2};\\
    &\le 2||A||_2^2\left(\sum_{m=1}^{d_0+d_1}\left(|| U\proj{m}(U^\dagger -V^\dagger)||_2 - ||(U- V) \proj{m}V^\dagger||_2\right)^2\right)^\frac{1}{2};\\
    &\le 2||A||_2^2\left(\sum_{m=1}^{d_0+d_1} \left(|| U\proj{m}||_2+|| \proj{m}V^\dagger||_2\right)^2||U-V||^2_2 \right)^\frac{1}{2};\\
    &= 4 \sqrt{d_0+d_1} ||A||_2^2 ||U-V||_2.
\end{align*}
\twocolumngrid
In the above evaluation we made elementary use of triangle inequality, Cauchy-Schwartz inequality, Hölder's matrix inequality, and also employed the definition of $||\cdot||_2$ and its sub-additivity property.  With all that, and using $||0\oplus \idty/d_0||_2=||\idty/d_0\oplus 0||_2=d_0^{-1/2}$ we can established a Lipschitz constant for $\tr[(\rho_\infty^H)^2]$ equal to $2 \sqrt{d_0+d_1}/d_0$.  Finally, employing the union bound, we reach the desired result shown in \eqref{eq:levy_dist}.

For the bound on the distance we also must show  that $\sigma$, in \eqref{eq:bound_dist}, vanishes for large apparatuses. This is done as follows. First we write $\sigma = \overline{\tr[(\rho_\infty^H)^2]}-\tr[(\overline{\rho_\infty})^2]$. The second term in this expression can be directly evaluated from \eqref{eq:fully_avg_state}, and reads:
\begin{multline}
        \tr[(\overline{\rho_\infty})^2]= (|c_+|^4+|c_-|^4)\left[ \left(\frac{d_1}{d_0+d_1+1}\right)^2\frac{1}{d_1} + \right. \\ \left. +\left(\frac{d_0+1}{d_0+d_1+1}\right)^2\frac{1}{d_0} \right].
        \label{eq_app:typycal_purity}
\end{multline}
For the first term of $\sigma$ we write:
\begin{multline*}
    \overline{\tr[(\rho_\infty^H)^2]}=\\ |c_+|^4\tr\left\{ \sum_{m=1}^{d_0+d_1} \overline{\left[ P_m^+\left(0\oplus\frac{\idty}{d_0}\right)P_m^+\right]}\left(0\oplus\frac{\idty}{d_0}\right)\right\} +\\
   +|c_-|^4\tr \left\{\sum_{m=1}^{d_0+d_1} \overline{\left[ P_m^-\left(\frac{\idty}{d_0}\oplus 0\right)P_m^-\right]}\left(\frac{\idty}{d_0}\oplus 0\right)\right\}.
\end{multline*}
Using standard results on Haar integrals \cite{Mele2024} to perform the average, we get:
\begin{equation}
\overline{\tr[(\rho_\infty^H)^2]}= \frac{|c_+|^4+|c_-|^4}{d_0+d_1+1} \left(\frac{1}{d_0}+1\right).
\label{eq_app:avg_purity}
\end{equation}
Performing the difference between the above expressions, and using that $|c_+|^4+|c_-|^4\le 1$, we reach the result written in \eqref{eq:bound_dist}.

\end{document}